\documentclass[aps,pre,twocolumn,showpacs]{revtex4}
\usepackage{epsfig}
\usepackage{times}
\begin{document}

\title{Congestion phenomena caused by matching pennies in evolutionary games}

\author{Gy\"orgy Szab\'o$^{1,2}$ and Attila Szolnoki$^1$}
\affiliation{$^1$ Institute of Technical Physics and Materials Science, Research Centre for Natural Sciences, Hungarian Academy of Sciences, P.O. Box 49, H-1525 Budapest, Hungary \\
$^2$ Regional Knowledge Centre, E{\"o}tv{\"o}s University, Ir{\'a}nyi D{\'a}niel u. 4, H-8000 Sz{\'e}kesfeh{\'e}rv{\'a}r, Hungary \\
}

\begin{abstract}
Evolutionary social dilemma games are extended by an additional matching-pennies game that modifies the collected payoffs. In a spatial version players are distributed on a square lattice and interact with their neighbors. Firstly, we show that the matching-pennies game can be considered as the microscopic force of the Red Queen effect that breaks the detailed balance and induces eddies in the microscopic probability currents if the strategy update is analogous to the Glauber dynamics for the kinetic Ising models. The resulting loops in probability current breaks symmetry between the chessboard-like arrangements of strategies via a bottleneck effect occurring along the four-edge loops in the microscopic states. The impact of this congestion is analogous to the application of a staggered magnetic field in the Ising model, that is, the order-disorder critical transition is wiped out by noise. It is illustrated that the congestion induced symmetry breaking can be beneficial for the whole community within a certain region of parameters.
\end{abstract}


\maketitle

\section{Introduction}
\label{sec:introduction}

Multi-agent evolutionary games on networks \cite{maynard_82, weibull_95, nowak_06, sigmund_10} are frequently used approach to study evolutionary processes characteristic to biological and social systems. In these models players are located on the sites of a network (or lattice) and their strategies represent biological entities or individual choices. For most of the cases the pair interactions among players are characterized by two-player games \cite{neumann_44}.

In these evolutionary games a uniform, symmetric, two-player, two-strategy game defines the payoffs for both players \cite{szabo_pr07}. In that simple cases we can distinguish only four possible strategy pairs (henceforth strategy profiles) for the elementary interactions.
\begin{figure}[ht]
\centerline{\epsfig{file=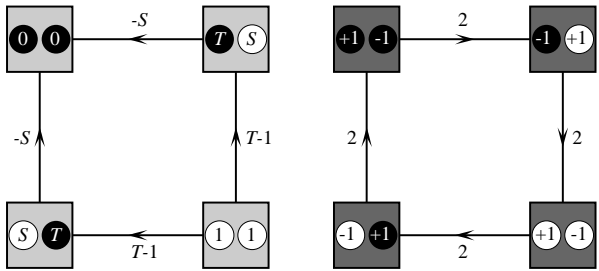,width=8cm}}
\caption{Flow diagrams for two-payer two-strategy games. The left and right diagrams show the payoffs for the prisoner's dilemmas ($S<0$ and $T>1$) and matching-pennies game. Arrows along the edges point towards the preferred strategy profile from the view point of player who modifies unilaterally her strategy.} \label{fig:flowdiag}
\end{figure}
In Fig.~\ref{fig:flowdiag} these strategy profiles are illustrated by large boxes with white and black bullets representing the two possible strategies for each player. The actual payoffs are denoted within these bullets. For the prisoner's dilemma, as the most challenging social dilemmas, the black and white bullets refer to Defection ($D$) and Cooperation ($C$) strategies while for the matching-pennies game these symbols denote Head and Tail (two sides of a coin) states. Henceforth we use the abbreviated nomination of $C$ and $D$ strategies and the traditional notation of payoffs introduced for the investigation of social dilemmas \cite{macy_pnas02,santos_pnas06}.

The mentioned games represent two significantly different behaviors as it is illustrated in the flow diagrams where the edges connect those strategy profiles that can be transformed into each other by a single strategy change of a player. For the symmetric two-player, two-strategy games the payoff variations of the active player from both the $DD$ and $CC$ strategy profiles are equivalent. In fact, this is the reason why the sum of the payoff variations of active players along the single loop is zero independently of the values of $S$ and $T$. Consequently, all the symmetric two-player two-strategy games are potential games \cite{Monderer_geb96, monderer_jet96, szabo_pre14b}. For the potential games one can find a potential (as a function of strategy profile) summarizing the driving force for the unilateral strategy changes on the analogy of the potential energy in physical systems. For the multi-strategy systems potential can exist if the above condition is satisfied along all the possible loops in the space of strategy profiles. The existence of potential is accompanied with some general features. Namely, the maximal value of potential is achieved at a pure Nash equilibrium; additional pure Nash equilibria are represented by nodes without outgoing edges; and the absence of directed loops are forbidden in the flow diagram. For the potential games the latter feature results in that
one can arrive to one of the pure Nash equilibria when following the direction of arrows from any initial state for the multi-strategy and/or multi-agent games \cite{brown_gw_51, voorneveld_el00}.

If a multi-agent system is built up from two-player potential games then the potential for the whole system will be the sum of the pair potentials. Blume \cite{blume_l_geb93, blume_l_geb03} has shown that the evolutionary potential games evolve into a Gibbs ensemble if the evolution is controlled by a logit rule \cite{berkson_jasa44} that is similar to the Glauber \cite{glauber_jmp63} or Metropolis \cite{metropolis_jcp53} dynamics introduced for the investigation of kinetic Ising models. In other words, all these multi-agent systems with symmetric two-strategy interactions are equivalent to an Ising model \cite{herz_jtb94, szabo_pr07, galam_pa10, szabo_pre14}.

On the contrary, the flow diagram of matching-pennies game represents a fundamentally
different interaction. In that case the players are not equivalent and one of the players is always motivated to change her strategy. As a result, there is a uniform payoff increase (driving force) along the four edges that represents a directed loop in the flow diagram. For this game potential does not exist and the single Nash equilibrium is a mixed strategy profile where both players choose their strategy at random \cite{morris_94, cressman_03}. The matching-pennies game can be interpreted as a driving force creating circular transitions in the flow diagram and probability current loops in the dynamical graphs that can be quantified by evaluating the entropy production \cite{schnakenberg_rmp76, szabo_pre14}. The relevance of a conceptually similar circular transitions in biological systems was first described by van Valen \cite{van_valen_et73, van_valen_et80} who named this effect as Red Queen mechanism.
Similar endless races can be observed in social systems between buyers and sellers \cite{friedman_d_e91}, property owners and criminals \cite{cressman_cje98}, males and females \cite{cremer_epjb08, traulsen_prl05}, or conformists and rebels \cite{cao_zg_tcs14}. The consideration of an evolutionary multi-agent system on network with this type of interactions prescribes two kinds of players distributed on bipartite graphs or lattices \cite{xu_b_epjb14}.

The evolutionary matching-pennies game and its combination with an anti-coordination game were considered previously for several structured populations \cite{szabo_pre14}.
In the present paper we extend these analysis to systems where the mentioned probability current loops are distorted by bottleneck effects  which can modify the macroscopic behavior
significantly.

\section{The model}
\label{sec:model}

Now we study spatial evolutionary games with players located on a square lattice with periodic boundary conditions. Each player has two strategies and collects income with playing games with the four nearest neighbors. The pair interaction will be composed of a symmetric two-player two-strategy game (a social dilemma with $T-S$ parametrization) and a matching-pennies game. Our analysis will focus onto the region of hawk-dove games ($T>1$ and $S>0$) where chessboard-like strategy arrangements occur for myopic or Glauber type strategy updates preferred in statistical physics \cite{sysiaho_epjb05, roca_epjb09, szabo_pre14}.

Both the suitable description of the chessboard-like strategy arrangements and the application of the matchingpennies game require the division of the square lattice into two equivalent sublattices denoted as $\alpha=X$ or $Y$ resembling the black and white squares on a chessboard.
 For this sublattice division each site $x$ ($x \in X$) is surrounded by four nearest neighbors belonging to the opposite sublattice ($x+\delta \in Y$) and {\it vice versa}. It is convenient to denote the strategies of players at sites $x$ and $y$ by unite vectors as
\begin{equation}\label{eq:purestrats}
{\bf s}_x, {\bf s}_y= D = \left( \matrix{1 \cr 0 \cr}\right)\,,\mbox{ or } \,\,
         C=\left( \matrix{0 \cr 1 \cr}\right)\,.
\end{equation}
Using this traditional notation the player's income can be given by a sum of matrix products
\begin{equation}\label{eq:payoff}
u_x= \sum_{\delta} {\bf s}_x \cdot {\bf A} {\bf s}_{x+\delta} \,,\,\,\,\,
u_y= \sum_{\delta} {\bf s}_y \cdot {\bf B} {\bf s}_{y+\delta} \,,
\end{equation}
where the summation runs over the four nearest neighbor sites belonging to the opposite sublattice. According to the model definition the pair interactions of nearest neighbors are the composition of a social dilemma and a cyclic game. The  corresponding payoff matrix is given as:
\begin{equation}\label{eq:poms}
{\bf A}=\left(\matrix{ \varepsilon & T-\varepsilon \cr
                       S-\varepsilon & 1+\varepsilon \cr} \right) \,,\,\,\,\,\,
{\bf B}=\left(\matrix{ -\varepsilon & T+\varepsilon \cr
                       S+\varepsilon & 1-\varepsilon \cr} \right) \,,
\end{equation}
where $\varepsilon$ quantifies the strength of matching-pennies component.

During the elementary step of the evolutionary process a player is chosen randomly and she modifies her strategy from ${\bf s}_x$ to ${\bf s}_x^{\prime}$ with a probability depending on the payoff difference ($u_x^{\prime}-u_x$) between the final and initial states. Namely, the strategy update probability is given as
\begin{equation}
W({\bf s}_x \to {\bf s}_x^{\prime}, {\bf s}_{-x}) =\frac{1}{1+\exp[(u_x-u_x^{\prime})/K]} \,,
\label{eq:dyn}
\end{equation}
where the strategy profile for the rest of players, denoted traditionally as ${\bf s}_{-x}$, remains unchanged.

This transition favors the state providing higher payoff for the player $x$ and similar rule is applied for the players of the opposite sublattice. Notice that $W({\bf s}_x \to {\bf s}_x^{\prime}, {\bf s}_{-x}) \simeq 1$ (or 0) if $u_x \ll u_x^{\prime}$ (or $u_x \gg u_x^{\prime}$) while the ''width'' of the transient region is proportional to $K$ that measures the magnitude of noise or the strength of selection. Noteworthy that the probability of strategy change [given by Eq. (\ref{eq:dyn})] is equivalent to those suggested by Glauber \cite{glauber_jmp63} for the kinetic Ising model.

For $\varepsilon = 0$ the above system satisfies the condition of potential games \cite{Monderer_geb96} because the sum of payoff variation along the single loop is zero, as illustrated in Fig.~\ref{fig:flowdiag}. In fact the kinetic Ising model and the corresponding two-strategy evolutionary potential games are equivalent, thus the coupling constant ($J$) between the neighboring spins ($s_x=\pm 1$) and the strength ($h$) of the external magnetic field can be expressed by the payoff parameters as $J=(1-S-T)/4$ and $h=1+S-T$ while $K$ corresponds to the temperature in the thermodynamical system \cite{herz_jtb94, szabo_pr07, galam_pa10, szabo_pre14b}. The effect of $\varepsilon$ on the transition from the anti-ferromagnetic to the paramagnetic phase as a function of $K$ is investigated in a recent paper for the absence of external field $h$ \cite{szabo_pre14}. It is found that the disturbance of the matching-pennies component reduces the critical value of $K$ while the universal (Ising type) features of the critical transition are preserved.

In the present work we explore how the stationary states of an evolutionary social dilemmas are influenced by the matching-pennies game in the presence of an external field ($h \ne 0$) representing games with $S \ne T-1$. In the literature of Ising models \cite{domb_74} it is well described that at zero temperature the ordered anti-ferromagnetic state is transformed into the preferred ferromagnetic state if $h$ exceeds a threshold value proportional to $J$ (the pre-factor depends on the number neighbors determined by the lattice structure). On the other hand, by increasing temperature $K$ the system exhibits an Ising type critical phase transition from the ordered anti-ferromagnetic state to the disordered spin arrangement at a critical point ($K_c$) decreasing with $|h|$ \cite{griffiths_prl70, rapaport_jpc71, penney_pa03}. Now it will be shown that the latter robust critical transition is smoothed out for $|\varepsilon| > 0$. For this purpose we first study a two-player game exhibiting the congestion phenomenon that results in macroscopic changes in the multi-player systems.

\section{Congestion for two-player system}
\label{sec:twoplayer}

\begin{figure}[ht]
\centerline{\epsfig{file=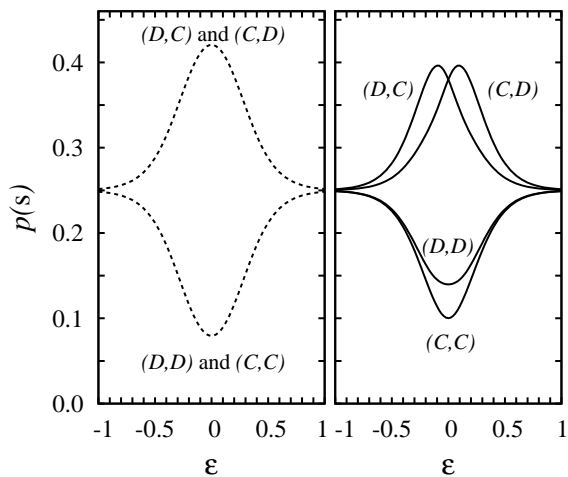,width=7.6cm}}
\caption{Probabilities of strategy pairs as a function of $\varepsilon$ for $T=1.5$, $S=0.5$ (left panel, dashed lines) and $T=1.4$, $S=0.3$ (right panel, solid lines) at $K=0.3$.}
\label{fig:probs}
\end{figure}

In this section we study a two player game with the same dynamical rule introduced in the previous section. In that case the system has only four microscopic states [${\bf s}=(D,D)$, $(D,C)$, $(C,D)$, or $(C,C)$] and the corresponding probabilities $p({\bf s})$ in the stationary state can be determined numerically by solving the suitable equations of motion for any values of payoff parameters and noise levels. In the absence of neighborhood, we can apply a simplified version of the traditional pair approximation (for a brief survey see \cite{hauert_ajp05, szabo_pr07}). Figure \ref{fig:probs} compares the configuration probabilities in dependence of $\varepsilon$ for the all the four possible strategy pairs at a fixed noise level.
The dashed lines in the left panel of Fig.~\ref{fig:probs} represent the behavior when the $(D,D)$ and $(C,C)$ strategy pairs appear with the same probability for arbitrary values of $\varepsilon$.
This is equivalent to $h=0$ case for the terminology of Ising model. Here $p(D,C)=p(C,D)$ for any values of $\varepsilon$ and all the four configuration probabilities tend to $1/4$ if the matching-pennies components dominate the game, as it is discussed in \cite{szabo_pre14}. The latter behavior reflects the fact that the matching-pennies game has only one mixed Nash equilibrium when both players choose their strategies at random. Notice furthermore that the variations in the configuration probabilities are even functions of $\varepsilon$.

\begin{figure}[ht]
\centerline{\epsfig{file=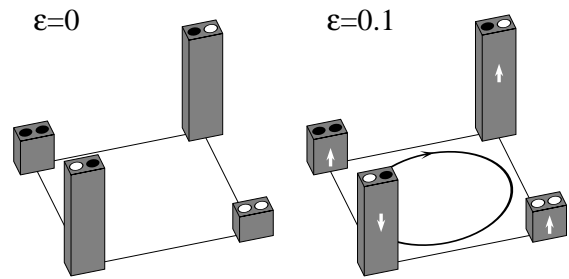,width=7.6cm}}
\caption{Configuration probabilities are illustrated by the height of columns for the absence (left plot) and presence (right plot) of matching-pennies component for $T=1.4$ and $S=0.3$. In the right plot the directed ellipse illustrates the emergence of a probability current loop that modifies the configuration probabilities via a bottleneck effect and the white arrows on the faces of columns show how the probabilities change due to the congestion phenomenon.}
\label{fig:congestion}
\end{figure}

Significantly different behavior can be observed if $1+S-T=h \ne 0$ even for $\varepsilon=0$. The right plot of Fig.~\ref{fig:probs} shows a situation when the preferred $(C,D)$ and $(D,C)$ strategies are present with the same high probability while $p(D,D) > p(C,C)$ that is obtained for the payoff parameters given in the caption of Fig.~\ref{fig:probs}. It is emphasized that in the absence of matching-pennies component the system satisfies the condition of detailed balance, that is, the forward and backward transitions appear with the same probabilities along the four edges of the flow diagram as it is illustrated in the left plot of Fig.~\ref{fig:congestion}.

The presence of the matching-pennies component, however, breaks the detailed balance and induces a probability current through the four-edge loop with a strength characterized by the difference of transition frequencies between the forward and backward directions along the four edges.
In the steady state these probability currents should be equivalent due to Kirchhoff laws \cite{kirchhoff_apc1847}. This uniform probability current modifies the configuration probabilities $p({\bf s})$ and destroys the equivalence between $(D,C)$ and $(C,D)$ configurations because the $(D,D)$ and $(C,C)$ states represent different barriers (widths of bottleneck) for the probability current flow. As a result, $p(D,C)$ [$p(C,D)$] increases [decreases] linearly with with the strength of the matching-pennies component for small values of $\varepsilon$. The white arrows in Fig.~\ref{fig:congestion} show that at the bottleneck states the corresponding probabilities increase as it happens before the narrower bottleneck while the probability of the fourth should be reduced because $\sum_{\bf s} p({\bf s})=1$.

Notice that the linear dependence on $\varepsilon$ implies that the behaviors of $p(D,C)$ and $p(C,D)$ are exchanged when reversing the sign of $\varepsilon$, that is accompanied with a reversal of probability current, too. For high values of $\varepsilon$ the system tends towards a uniform strategy distribution as mentioned above. Consequently, if $\varepsilon$ is increased then $p(C,D)$ exhibits a local maximum at $\varepsilon \simeq 0.1$ if $T=1.4$, $S=0.3$, and $K=0.3$. For the opposite sign of $\varepsilon$ the variations of $p(D,C)$ and $p(C,D)$ are exchanged as shown in Fig.~\ref{fig:probs}. The results of this congestion phenomenon is resembling the impact
of a staggered magnetic field $h_s$ ($h_s$ is positive for the sites of sublattice $X$ and negative on the sublattice $Y$) for the anti-ferromagnetic Ising models that prefers one of the ordered arrangements to the other.

The driving effect of the matching-pennies component seems to be similar to the effect of external electric field in driven lattice gas models \cite{katz_jsp84}. For these models the external field drives the system out off the Boltzmann distribution \cite{schmittmann_95} and generates a permanent particle transport when periodic boundary conditions are applied. It breaks inherent symmetries \cite{szabo_pra91}, enhances the effect of noise and the impact of local inhomogeneities \cite{szabo_pre94}. In case of repulsive interaction, where similar checkerboard-like ordered phase is formed in the half-filled lattice \cite{katz_prb83, dickman_pra90}, the order-disorder phase transition is modified due to the particle transport \cite{szabo_pre97}. Lastly we note that the effect of bottlenecks along a ring was recently studied by considering the asymmetric exclusion process \cite{sarkar_pre14} that is the simplest version of the driven lattice gases. The latter investigation is motivated by biological experiments indicating unidirectional circular ribosome translocations along messenger RNA loops with defects or slow codons in cells \cite{chou_bpj03}.

The application of the matching-pennies component in evolutionary games induces local probability currents within many significantly shorter loops that can be quantified even for two neighboring players when considering the transitions among the four possible strategy profiles for most of their quenched neighborhood \cite{szabo_pre14}.

\section{Congestion for multi-player system}
\label{sec:multiplayer}

\begin{figure}[ht]
\centerline{\epsfig{file=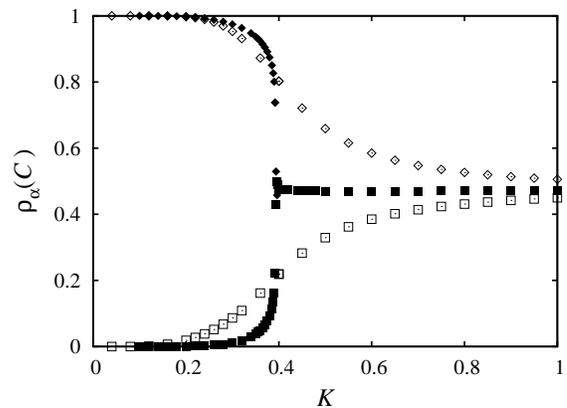,width=7.5cm}}
\caption{Frequency of strategy $C$ as a function of $K$ in sublattice $X$ (open diamonds) and $Y$ (open boxes) for $T=1.4$, $S=0.3$, and $\varepsilon=0.1$. The filled symbols indicate the same quantities for $\varepsilon = 0$.}
\label{fig:sop_k}
\end{figure}

The above-mentioned weak effect of the matching-pennies component remains valid for each pair and it is amplified for the multi-agent games. As a result the macroscopic system will evolve into the preferred ordered strategy distribution at sufficiently low noise levels. This fact is illustrated in Fig.~\ref{fig:sop_k} where the frequency of strategy $C$ ($\rho_{\alpha}(C)$) is plotted as a function of noise level $K$ for both sublattices ($\alpha = X, Y$). For $\varepsilon = 0$ this system shows an Ising type critical transition at $K_c=0.3922(1)$ from the sublattice ordered strategy arrangement to a disordered state \cite{szabo_pre14}. This critical transition, however, is smoothed out if the game includes the matching-pennies component.

Open symbols in Fig.~\ref{fig:sop_k} show the Monte Carlo (MC) results we obtained for $\varepsilon = 0.1$ on square lattice with a linear size $L=400$. In that cases the sampling and relaxation times are chosen to be $t_s=t_r=10,000$ MCS where within the time unit (MCS) each player has a chance once on average to modify her strategy. When reversing the sign of $\varepsilon$ then the strategy occupations in the sublattices are exchanged. As we argued, this phenomenon is a straightforward consequence of the congestion effect we described in the two-player model. Furthermore, the sublattice occupations tends toward the prediction of Ising model if $\varepsilon \to 0$ in a way resembling the vanishing of the staggered magnetic field. Evidently, we have to use significantly larger system sizes ($L=2000$) and longer runs ($t_s=t_r=10^6$ MCS) for obtaining the reference data at $\varepsilon = 0$ in order to suppress the undesired effects of the diverging fluctuations and critical slowing down when approaching the above-mentioned critical point.

\begin{figure}[ht]
\centerline{\epsfig{file=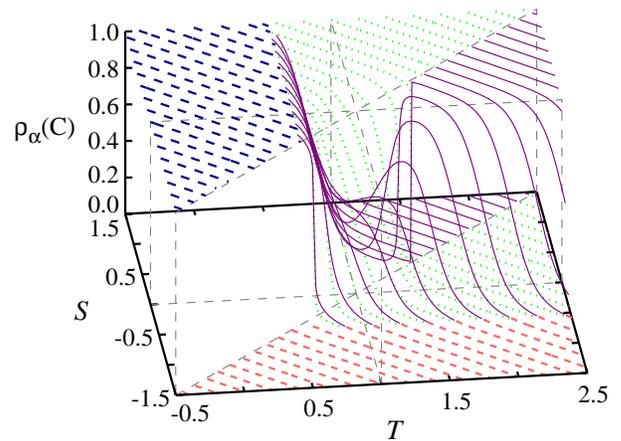,width=8.0cm}}
\caption{(color online) Frequency of strategy $C$ as a function of payoff parameters $T$ and $S$ in both sublattices for $K=0.3$ and $\varepsilon=0.1$. Dashed (red and blue) lines illustrate MC data where the sublattices are equivalent. The solid (purple) and dotted (green) lines show $\rho_C$ in the sublattices $X$ and $Y$.}
\label{fig:sop_TS}
\end{figure}

 Our numerical analysis is repeated in the full payoff parameter region, that covers all social dilemma games at fixed values of $\varepsilon$ and noise level ($K=0.3$). The MC data are summed in Fig.~\ref{fig:sop_TS} where the lines are obtained by varying $T-S$ at fixed values of $T+S$.
The MC results indicate the absence of sublattice ordering, that is $\rho_X(C)=\rho_Y(C)$, within a large regions of the prisoner's dilemma, harmony, and stag hunt games. The $C$ strategies dominate equally both sublattices in the zero noise limit if $T<1$ and $(T-1-S)<0$ and the homogeneous $D$ state is stable if $S<0$ and $(T-1-S)>0$ for the zero noise limit ($K \to 0$).

Staying at the same limit only one of the sublattice ordered structures appears in the hawk-dove game quadrant ($T>1$ and $S>0$). More precisely, $\rho_X(C) \to 1$ and $\rho_Y(C) \to 0$ if $\varepsilon>0$ and $(T-1-S)<0$ or when $\varepsilon<0$ and $(T-1-S)>0$. Evidently, when reversing the sign of $\varepsilon$ and $(T-1-S)$ separately then the preference between the two sublattice ordered structures are exchanged.

The similarity between the impact of staggered magnetic field in the Ising model and the introduction of a matching-pennies component in the hawk-dove game can be demonstrated by monitoring the motion of an interface separating the two sublattice ordered arrangements. For both cases the visualization of the evolutionary process shows clearly the expansion of the preferred sublattice ordered structure via the motion of separating interface.

It is emphasized that the two-fold degeneracy can be observed only along the line $T-1-S=h=0$ separating regions where the chessboard and anti-chessboard arrangements of the $D$ and $C$ strategies occur as it is shown in Fig.~\ref{fig:sop_TS}. Noteworthy, that Ising type critical transitions can take place only along this line if the noise level is increased.

When decreasing the values of $\varepsilon$ and $K$ then the widths of the intermediate
transition regions shrink, too. Additionally, one can observe a disordered state at close vicinity of the central point ($T-1=S=0$) of the $T-S$ plane where both the coupling constant $J$ and the external magnetic field $h$ vanish in the equivalent Ising model. Evidently, the absence of ordering will also characterize the system behavior if $\varepsilon$ and/or $K$ exceed threshold values depending on $T$ and $S$.

\section{Exploitation of congestion}
\label{sec:exploitation}

The systematic analysis of the evolutionary games has highlighted the existence of many different mechanisms enhancing the total income of societies even for the cases of social dilemmas (for a survey see \cite{nowak_s06, szabo_pr07, sigmund_10, perc_bs10}). It is already known that the total income of society can be influenced by the game itself (payoffs and strategy set), the connectivity network, the dynamical rules including the noise level and allowing co-evolutionary processes in all ingredients of the systems. In the light of these results the utilization of congestion phenomenon emerges directly. The relevance of this question is stressed by recent experiments investigating human behavior in real-life situations involving the matching-pennies component in the payoffs \cite{belot_pnas13, cao_zg_pone13, cao_zg_tcs14}.

The above results justify that the main impact of matching-pennies is related to the preference of one of the chessboard-like strategy arrangements that suppresses the critical transition related to the existence of two equivalent (optimal) ordered strategy arrangements. As a result, the preferred structure dominates the system behavior if $K > K_c$ within the region of hawk-dove games. The latter effect can increase the total income because of the reduction of the length of interfaces responsible to the loss in the total payoff. Conceptually similar congestion induced increase of collective income may be expected in human societies.

For an illustration of the mentioned phenomenon we compare the average payoffs as a function of noise $K$ for three different evolutionary rules whereas the payoff parameters ($T$ and $S$) are fixed. For the selected payoffs ($T=1.9$ and $S=0.3$) the maximum average payoff [$2(T+S)$] is achieved if the $C$ and $D$ strategies form a chessboard-like structure in the zero noise limit. It is emphasized that the matching-pennies component does not affect the average payoff as it is a zero-sum game.

\begin{figure}[ht]
\centerline{\epsfig{file=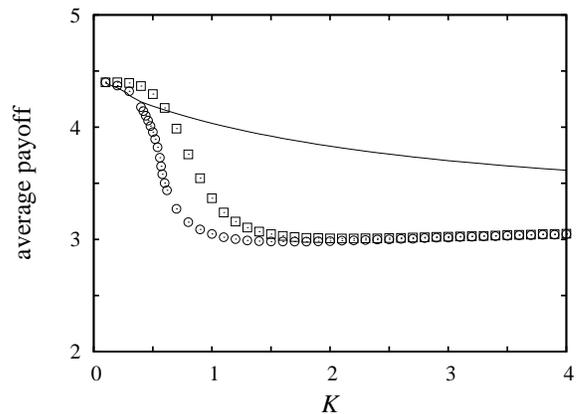,width=7.8cm}}
\caption{Comparison of average payoffs versus noise ($K$) for evolutionary games on square lattice at $T=1.9$ and $S=0.3$. Open circles and boxes represent MC data obtained for $\varepsilon = 0$ and $0.1$ if the evolution is controlled by the logit rule defined by Eq. (\ref{eq:dyn}). Solid line illustrates MC results obtained when the evolution is controlled by collective pairwise strategy update \cite{szabo_pre10}.}
\label{fig:pocomp}
\end{figure}

Figure \ref{fig:pocomp} illustrates that the presence of matching-pennies component increases the average payoff for the given payoff parameters. The most relevant increase occurs in the vicinity of the critical noise level [$K_c(T=1.9,S=0.3) = 0.590(4)$].
The importance of congestion phenomenon becomes more striking when the average payoff is compared to results obtained for other types of dynamical rules. Within the hawk-dove region the imitation-based rules can not be considered as an adequate reference because the imitation of the neighboring strategy prevents the formation of the optimal strategy arrangement \cite{hauert_n04}.  Up to now one of the highest average payoffs is achieved by the application of the collective pair-wise strategy update \cite{szabo_pre10} where the stochasticity is introduced via a noise parameter $K$, too. For the latter rule the fraternal players take into consideration the co-player's income. The corresponding MC data of the average payoffs are illustrated by the solid line in Fig. \ref{fig:pocomp}. Accordingly, there is a region of $K$ where the introduction of the
matching-pennies payoffs provide the highest average income for the whole population.

\section{Summary}
\label{sec:summary}

In this paper we have studied two-strategy evolutionary games on square lattice when matching-pennies games with a strength $\varepsilon$ are added to the payoffs in order to modify the interactions between neighboring players. Here matching-pennies game represents the simplest cyclic dominant game for the description of Red Queen effect in two-strategy games if the evolutionary process is controlled by the logit rule resembling the Glauber dynamics in statistical physics. In that case the matching-pennies component of the pair interactions can be considered as a microscopic driving force that destroys the detailed balance by inducing probability current loops throughout the four microscopic states for each pair interactions.

The presence of the matching-pennies component eliminates the equivalence between the players residing in the two sublattices. Additionally, the equivalence between the two ordered (chessboard- and anti-chessboard-like) strategy distributions is also destroyed via a contagion mechanism affecting the configuration probabilities if the disfavored strategy pairs [{\it e.g.} $(D,D)$ and $(C,C)$] would be present with different probabilities at $\varepsilon=0$. The direction of preference varies with the sign of $\varepsilon$. Note that the presence of both ordered phases with equal weights is disadvantageous because along the separating interfaces frustrating players cannot enjoy sufficiently high payoff. This ambiguity can be resolved by matching-pennies component in a desired way.

Finally we emphasize that similar cyclic dominance can be created by the rock-paper-scissors component in three-strategy evolutionary games for equivalent players as it is reported for the description of many systems (for a survey see refs. \cite{nowak_06, szabo_pr07, sigmund_10, szabo_pre14b}). In the latter cases, however, the matching-pennies components and their effects can be recognized within all the $2 \times 2$ sub-games where the interacting players are limited to use two different strategy pairs. The latter feature underlines the importance of the matching pennies games as it can be included within the $2 \times 2$ sub-games for most of the symmetric $n \times n$ ($n>2$) matrix games.

\section*{Acknowledgments}

This work was supported by the John Templeton Foundation (FQEB Grant \#RFP-12-22), the Hungarian National Research Fund (OTKA TK-101490), and the European Social Fund through project FutureICT.hu (TAMOP-4.2.2.C-11/1/KONV-2012-0013).


\end{document}